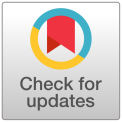
Check for updates

# Development of Cybersecurity Norms for Space Systems

Samuel Sanders Visner[1] (AIAA Number 987385)
Peter Sharfman, Ph.D.[2]
*The MITRE Corporation, McLean, VA 22102 USA*

**This paper addresses:**

- **Evolution of the space systems environment, including space system proliferation and space systems as critical infrastructure**
- **Cyber threats to, and vulnerabilities of, space systems**
- **Alternative approaches to meeting these threats, and the significance of norms**
- **Approaches to the development and reinforcement of norms for the cybersecurity of space systems.**

## 1. The Space Systems Environment

The current and forecast proliferation of space systems constitutes a fundamental change in the space systems environment. Over 3,000 satellites are now active. In addition to government launch services in several countries, many space systems are being developed and operated by the private sector. One operator alone, Starlink, already launches 60 satellites at a time and plans to build a global 5G network supported by 12,000 low earth orbit satellites. According to Bank of America,[3] the global economic value of the space industry may triple to $1.4 trillion by 2030, by which time an estimated 50,000 satellites could be in orbit, many connected to public communication networks and global cloud infrastructures.

Government-owned and -operated space systems also continue to expand. China is today a major spacefaring country, lofting satellites for resource mapping, national security, civilian communication, and exploration.[4] Missions by the United States, EU, Russia, Japan, Israel, Iran, and India demonstrate the importance of space to national interests and prestige.

For example, agricultural operations depend on and are made more efficient through the use of GPS-guided combine-harvesters and Geographic Information Systems. The US Government notes:

> *GPS equipment manufacturers have developed several tools to help farmers and agribusinesses become more productive and efficient in their precision farming activities. Today, many farmers use GPS-derived products to enhance operations in their farming businesses. Location information is collected by GPS receivers for mapping field boundaries, roads, irrigation systems, and problem areas in crops such as weeds or disease. The accuracy of GPS allows farmers to create farm maps with precise acreage for field*

---

*areas, road locations and distances between points of interest. GPS allows farmers to accurately navigate to specific locations in the field, year after year, to collect soil samples or monitor crop conditions.*[5]

Global 5G networks, connected to global cloud infrastructures, will serve global commerce and commercial operations. Media reports:

> *Microsoft Azure has announced a partnership with SpaceX that will give customers the ability to both access and deploy cloud computing capabilities anywhere on Earth with the help of Starlink internet.*[6]

Similar arrangements are reported between BlueOrigin's global 5G network and Amazon Web Services.

A recent analysis performed by the Space Information Sharing and Analysis Center (Space ISAC) of the dependence of National Critical Functions on space systems, makes evident the extent to which cyber (and physical) disruption of space systems would jeopardize the national economy, and the safety and well-being of people who depend on the nation's critical infrastructures.

Other reasons exist to safeguard space systems and to create norms associated with doing so. Cyber-attacks on space systems important to intelligence and national defense, particularly those associated with warning of strategic attack, could be destabilizing, and seen as presaging even more aggressive and dangerous developments.

Overall, space systems exist in a transnational environment, one in which each country's systems orbit over the territory of other, perhaps many other, countries. Disruption of space systems belonging to one country can affect communication, commerce, and navigation in many countries. While this may be true of other infrastructures, the transnational nature of space operations makes the cybersecurity of space systems a matter that should be of common interest to all countries.

A broad discussion is under way in the United States[7] to determine whether space systems should be considered "critical infrastructure," as defined by Presidential Policy Directive 21.[8] An Aspen Institute report notes such a designation "will facilitate prioritization of limited government resources, grow personal relationships between industry operators and policymakers, and help overcome obstacles to interagency coordination."[9] That report refers to Space Policy Directive 5 of the White House Space Council, which "places emphasis on the need to improve cyber protections when developing space systems, which are defined as a combination of a ground control network, a space vehicle, and a user or mission network that provide a space-based service."[10] The Space ISAC[11] has established a working group to examine if space systems should be declared critical infrastructure, and how such a declaration could be implemented. The Department of Homeland Security's Cybersecurity and Infrastructure Security Agency has established a working group to determine the manner and extent to which the National Critical Functions[12] depend on space systems.

In short, preserving the functionality of space systems, which has been of critical importance to the US military and intelligence capabilities for decades, has now become critically important for the US economy and society, and arguably is rapidly becoming critically important for the entire world.

## 2.  Cyber Threats to Space Systems

As the economic importance of space systems grows, they will become a more attractive target for cyber-attacks, while maintaining their cybersecurity will become increasingly important to US national and economic security and to the stability of the overall international system. Vulnerabilities may allow attackers to hack satellite control and navigation systems, spoof satellites into accepting unauthorized transmissions, or simply jam satellite communication. In 2019, the Defense Intelligence Agency[13] reported that Russia and China are "developing jamming and cyberspace capabilities, directed energy weapons, on-orbit capabilities, and ground-based anti-satellite missiles that can achieve a range of reversible to nonreversible effects."

---

[5] See: https://www.gps.gov/applications/agriculture/

[6] See: https://www.teslarati.com/spacex-starlink-microsoft-azure-partnership/

[7] https://www.nextgov.com/cybersecurity/2020/12/experts-call-government-designate-commercial-space-sector-critical-infrastructure/170439/

[8] https://obamawhitehouse.archives.gov/the-press-office/2013/02/12/presidential-policy-directive-critical-infrastructure-security-and-resil

[9] https://www.aspeninstitute.org/longform/a-national-cybersecurity-agenda-for-resilient-digital-infrastructure/securing-the-internets-public-core/

[10] https://aerospace.csis.org/how-does-space-policy-directive-5-change-cybersecurity-principles-for-space-systems/

[11] Co-author Samuel Sanders Visner is a member of the Space ISAC's Board of Directors.

[12] See: https://www.cisa.gov/national-critical-functions

Securing these systems is difficult and is likely to remain so for several reasons.

- Many space systems today are vulnerable to cyber-attack – real and significant – although some systems are more vulnerable than others.
- There is every reason to expect that hacking tools will continue to improve, which will increase the vulnerability of space systems.
- Most space systems derive their value from an ability to communicate with multiple end-users on the ground, which provides multiple avenues of potential attack
- Extensive efforts to create highly reliable technical solutions for cybersecurity (deterrence by denial) for terrestrial systems would so far failed to achieve this objective, and a technical cybersecurity solution for space systems would be even more difficult because physical access to a satellite in orbit is infeasible.

Threats to satellite systems are and will continue to be varied and serious, affecting the space segment (the satellites themselves), the ground segment ("elements of space systems and allows command, control and management of the satellite itself and the data arriving from the payload and delivered to the users"[14]), and the link segment, representing communication between the ground and satellites, and between the satellites themselves, both for command and control and for user communication.[15] These threats represent the potential of both computer network exploitation – used to steal information – and computer network attacks – used to damage information, the systems that process that information, and even the physical systems that rely on information and information systems.

Regarding exploitation, Oxford researcher James Pavur demonstrated at the 2020 Blackhat conference his ability to intercept, with equipment both inexpensive and widely available, a broad range of satellite communication, ranging from "*communications from a disabled tanker pulling into a port in Tunisia*" to *"(a) Polish real estate group's annual financial report."* In his words:

> *"From the biggest companies in the world to individuals in a coffee shop," he said, "people are leaking deeply sensitive information over satellite feeds. You can also see why someone would be tempted to leave their protocols unencrypted because the performance difference is so enormous."*[16]

Pavur notes also:

> *"I think that satellite companies are generally aware that the traffic they're sending is unencrypted but aren't necessarily aware of whether they have a legal obligation to protect that data."*[17]

Apart from exploitation that compromises the confidentiality of information, other, offensive cyber threats to satellite systems exist. GPS signals can be jammed, leading to erroneous navigation of planes, ships, cars, trucks, and other systems that require satellite-based navigation signals. Satellite uplinks and downlinks can also be jammed, thus denying users worldwide access to systems on which they rely increasingly. False command and control information can be sent to satellites, allowing attackers to alter satellite orbital dynamics. Of perhaps even greater significance, as satellite systems become an intrinsic component of global 5G networks, the ability of billions of Internet of Things devices to function correctly (or at all) may be put at risk.

---

[13]https://www.dia.mil/Portals/27/Documents/News/Military%20Power%20Publications/Space_Threat_V14_020119_sm.pdf
[14] Cyber Threats to Space Systems | Joint Air Power Competence Centre (japcc.org)
[15] Ibid.
[16] Hacking Cyber Space (forbes.com)
[17] The satellite-hacker's guide to the space industry: don't panic (yet) | CyberNews

Evidence is mounting that these threats are being realized. A 2018 Chatham House report noted Norwegian intelligence that Russia had disrupted NATO GPS signals during NATO's 2018 Trident Juncture exercise and that Russia is integrating this capability into its own operational concepts.[18]

## 3.  Alternative Approaches to Meeting the Threat

We can distinguish three broad approaches to countering a serious threat to any complex and high-value system:

- Hardening and defending the system in such a way that even a determined attack by the most capable potential adversaries is likely to fail. (This is sometimes described as "deterrence by denial.")
- Increasing the resilience of the system or creating alternatives to the system so that its essential functions continue even in the aftermath of a "successful" attack. (This is sometimes described as "attack mitigation.")
- Manipulating the perceptions and incentives of potential attackers so that they will choose not to attack the system. (This includes, but is not limited to, deterrence by threat of retaliation or punishment.)

In the case of potential cyber-attacks on space systems, the first of these approaches is likely to fail. The experience of the last few years with cyber-attacks on terrestrial systems is that any information system with extensive connections to other information systems is vulnerable to a capable and determined cyber-attack, even if the information itself is not exploited. It would probably be possible to protect a satellite from cyber-attack by designing it to communicate only with a limited set of ground stations, provided that the ground stations were themselves air-gapped from any other information systems, but such an arrangement would render the satellite almost useless.

The second of these approaches has been adopted to some extent by the US military and the US intelligence community. However, it is expensive, and for functions that involve high data rates it is prohibitively expensive.

This leaves as the primary avenue for action the manipulation of incentives. This in turn can be broken down into four alternatives:

- Creating a situation in which nobody has any reason to disrupt or deny US use of space systems. In a world of great power competition and extensive cybercrime, this can be ruled out for the foreseeable future.
- Creating a situation in which all potential attackers derive so much benefit from US space systems that their losses if these systems were disrupted or rendered inoperable would outweigh the gains they hoped to achieve. (This situation is sometimes referred to as "interdependence.")
- Creating a situation in which all potential attackers believe that if they launched a successful cyber-attack against a US space system, they would be subjected to a punishment that outweighed the gains they hoped to achieve. (This is the traditional meaning of "deterrence.")
- Creating and sustaining "norms" against cyber-attacks on space systems.

We argue that the latter three of these approaches can be made mutually reinforcing, and that in combination they offer meaningful protection to space systems. Before elaborating this argument, we need to explain what we mean by a "norm."

We define a "norm" as a rule of conduct generally recognized that is not a legal obligation or legal prohibition. Although the dividing line between a norm and a law is sometimes ambiguous, we can say in general that norms are less designed than laws to regulate behavior in detail; that unlike laws, norms have no established process for adjudicating claims of violation; and that unlike most laws, there is no formal mechanism for enforcement. (Enforcement is a slippery criterion, however – the Geneva Conventions are laws but cannot be enforced, while public opinion within a small community or a company can be a powerful enforcer of norms.) Norms, like laws, can continue to guide behavior even when they are occasionally violated, but cease to matter much when violations become pervasive and unpunished.

In many cases, norms determine the extent to which laws are obeyed. In the United States, there is a norm that automobile speed limits may be exceeded by a few miles per hour. While it is illegal both to steal sensitive military information and to assassinate political leaders, the prevailing norms dictate that international espionage takes place all the time, while international assassination is extremely rare.

---

[18] [2019-06-27-Space-Cybersecurity-2.pdf (chathamhouse.org)](https://chathamhouse.org)

Other things being equal, interdependence rests not only upon mutual or reciprocal benefits, but also upon expectations that all parties will continue to behave in ways that will allow these benefits to continue. Without such an expectation, there is a strong reason to hedge against a future breakdown of the interdependent relationship, and this hedging behavior will reduce the extent of the interdependence; often this hedging will create a feedback loop in which observation of hedging by others leads to hedging, which is in turn observed and treated as a reason for further hedging until the entire interdependent relationship breaks down. However, when interdependent behavior is sustained by expectations that it will continue, and the interdependent parties have generally similar expectations, these expectations become a norm.

Thus, interdependence creates norms, and norms enable interdependence to continue and sometimes to become deeper.

Similarly, norms and deterrence by threat of punishment are also mutually reinforcing. A threat to respond with punishment if a particular norm is violated has three advantages over a threat to punish some arbitrarily defined behavior. First, the party being threatened can assert that it is complying with a norm rather than complying with the demands of the threatener. Second, the party being threatened has less fear that compliance will encourage further threats and demands, since the precedent being set is one of norm compliance rather than successful bullying. Third, the party making the threat gains added credibility because it has a larger stake (or at least is perceived to have a larger stake) in the enforcement of norms than in the enforcement of some arbitrary demand. All three of these factors tend to increase the probability that the deterrent threat will succeed.

At the same time, a successful effort to use a threat of punishment to induce compliance with a norm has the additional effect of strengthening that norm for the future. The precedent of a successful deterrent threat to induce compliance with a norm creates or reinforces an expectation that any future attempt to violate the norm will be met with a response, and also the expectation that the response will be successful.

Thus, it appears that any effective strategy to prevent cyber-attacks on space systems will have to include the creation and/or the reinforcement of norms.

4.   Developing and Reinforcing Norms that Protect Space Systems from Cyber-Attack

While some limited cyber-attacks have likely occurred, as of the end of the summer of 2021, there has never been a publicly acknowledged cyber-attack against a space system. We can therefore say that a norm exists that such attacks should not take place – or at least that any cyber-attacks against a space system should be limited to those that will be kept secret not only by the attacker but also by the owner of the target. However, the previously mentioned trends toward proliferation of both space systems and cyber-attacks mean that this norm is fragile. Those who want to see continued growth in the benefits derived from the use of space and fear that these benefits may be jeopardized in the future by cyber-attacks therefore face a challenge: **how can the existing norm prohibiting cyber-attacks on space systems be clarified and reinforced?**

There are three ways to look at this norm, and those who seek to reinforce it can address it in any or all of these ways:

- Cyber-attacks on space systems as a variety of cyber-attack, and therefore norms on this subject as a subset or extension of norms regarding cyber-attacks
- Cyber-attacks on space systems as a variety of attack on space systems, and therefore norms on this subject as a variety or extension of norms regarding the treatment of space systems
- Cyber-attacks on space systems as a possibility to be isolated, with norms on this subject unrelated to norms regarding terrestrial cyber-attacks or kinetic attacks on space systems.

For some years, an effort has been under way to develop and codify norms to prohibit or limit cyber-attacks in general. The United Nations has convened a "Group of Governmental Experts work to deal with ICT threats in the context of international security" that has established five "pillars of work":

- Existing and emerging threats
- International law
- Norms, rules, and principles
- Confidence-building measures

- International cooperation and assistance in capacity-building.[19]

This effort, while making slow progress, has created a working group representing 193 UN-member countries, and issued a report[20] observing that voluntary, non-binding norms can contribute to conflict prevention. It called on states to avoid using information and communications technology "not in line with the norms for responsible State behavior." In other words, the violation of these norms – still to be defined – would lie outside the bounds of behavior considered acceptable by recognized states.

However, such norms would entail some drastic changes in existing patterns of behavior. As of the middle of 2021, the global pattern of behavior is:

- Cyber-attacks for the purpose of stealing money or stealing valuable intellectual property are widespread. Many governments tolerate such criminal activity by their residents provided that the target is outside their borders. While the US Government does not tolerate cyber theft by US-based criminals, it presumably engages in extensive theft of information for purposes of national intelligence and military intelligence. Cyber-attacks that cause disruption for the purpose of obtaining money (ransomware attacks) or coercing a private business (the North Korean attack on SONY) are treated in the same way.
- Cyber-attacks for the purpose of disrupting the normal functioning of information systems are infrequent and are treated as hostile activities short of outright war.
- In the aftermath of the 2021 ransomware attack on Colonial Pipeline, the United States has sought to create a norm that criminal activity that threatens critical infrastructure of the target country should not be tolerated by the country from which the attack was launched. It is too early to say whether this norm-creation effort will actually change future behavior.
- Several countries with authoritarian governments are attempting to create a norm prohibiting the use of cyber capabilities to compromise government control over the information available to their own citizens. The United States strongly opposes such norms.
- There is a norm prohibiting cyber-attacks aimed at killing people, such as by attacking an aircraft in flight or by surreptitiously altering medical records so that incorrect drugs are administered.

Given the existing patterns of behavior, and the absence of consensus among the countries with strong cyber capabilities over the desired content of norms regarding cyber-attacks, the protection of space systems by means of norms regarding cyber-attacks in general does not appear very promising.

In contrast, there is an extensive set of norms dealing with the protection of objects in space. The foundational document is the 1967 Outer Space Treaty,[21] which has been ratified by 111 countries.[22] Article 7 of this treaty holds states liable for any damage to persons or property of another state by an "object" launched into space. Article 6 makes each state responsible for activities carried out in space by its "non-governmental entities." While the Outer Space Treaty makes no mention of cyber-attack (a concept unknown when the treaty was drafted in 1966), it would not be much of a stretch to extend "liability" for kinetic attack (accidental or intentional) to prohibition of cyber-attack.

Furthermore, as we have already observed, existing patterns of behavior avoid the use of cyber to disrupt the functioning of space systems.

The Outer Space Treaty embodies the two principles that the United States has advanced in the ongoing discussions of cybersecurity norms without being able to forge a consensus around them:

- The touchstone for prohibited activity is damage to persons, objects, or economic activity, but not the dissemination of ideas or information that are unwelcome to some governments.

---

[19] https://cyberdiplomacy.disarmamenteducation.org/home/#first_section
[20] https://front.un-arm.org/wp-content/uploads/2021/03/Final-report-A-AC.290-2021-CRP.2.pdf
[21] Treaty on Principles Governing the Activities of States in the Exploration and Use of Outer Space, including the Moon and other Celestial Bodies, which entered into force on October 10, 1967.

[22] The only country with significant cyber capability that has not ratified the treaty is Iran, which has signed but not ratified.

- States are responsible for damage caused by persons under their jurisdiction.

It would be easier to get international agreement that these principles should remain in place in an existing treaty than to get agreement on a brand-new treaty that included these principles. Therefore, the option of treating cyber-attacks on space systems as an entirely new subject is tactically inferior to the option of extending the Outer Space Treaty to include a prohibition on cyber-attacks.

We conclude that the best approach is to extend the Outer Space Treaty's norm that prohibits kinetic attacks on space systems to prohibit cyber-attacks as well.

How might the US Government go about this? Three alternative strategies present themselves:

- Do nothing new, relying upon the fact that the existing pattern of behavior constitutes a norm prohibiting cyber-attacks on space systems.
- Make a public statement of policy that the United States believes that the principles embodied in the Outer Space Treaty should be understood to make states liable for any attack on another state's space system, whether that attack is kinetic or cyber. There is a range of possible formality, from a comment made at a presidential press conference to a Joint Resolution passed by both houses of Congress and signed by the President. Such a policy statement would be accompanied by encouraging other countries to issue similar policy statements.
- Propose amending Article 7 of the Outer Space Treaty to make it explicitly apply to cyber interference (and perhaps also to lasers or other directed energy) as well as to objects launched into space.

These alternatives are in ascending order of potential effectiveness in reinforcing the existing norm, but if a country known to have strong cyber capabilities explicitly dissented from the US policy, that would instead weaken the norm, and a failed effort to amend the Outer Space Treaty might weaken the norm still more. The US Government should carefully assess the likely reactions of other countries before settling on one of these strategies.

In any event, it will remain true that the norm is reinforced both by interdependence and by deterrence.

## 5. Finding and Recommendations

We find that cyber-attacks on space systems in peacetime damage our country's interests. Therefore, we view as beneficial the norm, currently observed though not codified formally, that such attacks do not take place generally. As a result, it is in the interests of the United States to support maintenance of this norm, while working to reinforce it through more formal mechanisms.

Our recommendations reflect this view of our interests. The United States should:

I. Encourage those who deploy commercial space systems to operate them in ways that benefit other countries as well as the United States and call these benefits to everybody's attention.

II. Respond to any cyber-attack on a US space system in a way that will strongly discourage any future attacks.

III. Engage in informal consultations with those countries that possess substantial space and/or cyber capabilities regarding a possible amendment to Article 7 of the Space Treaty extending its prohibitions on attacking space systems to the use of cyber weapons.

IV. If these consultations are not productive, issue a unilateral US policy statement that cyber-attacks on space systems are unacceptable.